\documentclass[twocolumn]{revtex4-1}
\usepackage{graphicx}
\begin{document}
\title{Physical Decomposition of Photon Angular Momentum}
\author{Wei-Min Sun$^{1,2}$}
\thanks{sunwm@chenwang.nju.edu.cn}
\address{$^1$ School of Physics, Nanjing University, Nanjing 210093, China}
\address{$^2$ Joint Center for Particle Nuclear Physics and Cosmology, Nanjing University and Purple Mountain Observatory, CAS, Nanjing 210093, China}

\begin{abstract}
We address the problem of angular momentum decomposition of a free photon.
We propose a natural and physical scheme for separating the total angular momentum operator for a free photon
into an orbital part and a spin part with the hope that it solves this long-standing problem and it could provide
a natural basis for further study of the problem of angular momentum decomposition of a gauge field system.
\end{abstract}

\maketitle

The photon, as the quantum of the electromagnetic field, is a basic constituent particle of the Standard Model. Like any other microscopic particle,
the photon can possess a certain angular momentum. However, for the photon, whether it is meaningful on physical ground to separate its total angular momentum into the contributions of an orbital part and a spin part is still a controversial issue. In standard textbooks of Quantum Electrodynamics (QED), for example, the book by Berestetskii {\it et al.}\cite{BLP}, which represents the viewpoint of Landau School, it is explicitly stated that one could only talk about the total angular momentum of a photon and the separation of the photon angular momentum is of no physical meaning. This viewpoint of Landau School is recognized as true by the physics community for many years.

In 2008, a paper by Chen, L$\rm{\ddot{u}}$, Sun, Wang and Goldman (hereafter referred to as Chen {\it et al.}) \cite{Chen} appeared in which the authors studied a problem related to this (and also more general than this), viz., that of separating the total angular momentum of a gauge field system (both QED and QCD) into the contributions of orbital and spin parts of the constituent matter particle (electron or quark) and gauge particle
(photon or gluon) fields. In that paper, by introducing a decomposition of the gauge potential into the so-called physical part and pure gauge part, the authors obtained a separation of the total angular momentum operator of QED and QCD in which each individual part is gauge-invariant, and, as the authors claim, also satisfies the angular momentum algebra. Chen {\it et al.}'s work evoked active responses of the community. After their work, some researchers further studied this approach. During these studies some different ways to perform the gauge potential decomposition were discovered and eventually people realized that there could be infinitely many different ways to perform such a decomposition. This implies that there are infinitely many different ways to split the angular momentum of a QED or QCD system in a gauge-invariant way! In this situation, one needs to answer which separation scheme is the physical one \cite{LL}. Besides, Chen {\it et al.}'s assertion that in their angular momentum decomposition scheme each individual part satisfies the angular momentum algebra has also been questioned by some researchers. Wakamatsu pointed out that in the simplest case of a free electromagnetic field the photon spin operator as defined {\it a la} Chen {\it et al.} does not satisfy the angular momentum algebra \cite{Wakamatsu}. In fact, the very question of whether one should demand that in a physically meaningful angular momentum decomposition scheme each part satisfy the angular momentum algebra is still controversial \cite{JiChen}. Up till now, all these questions have no definite answers yet. Therefore, with regard to the problem of separation of photon angular momentum, we could say that a separation of the total angular momentum of a photon into orbital and spin contributions is physically meaningful, but which separation scheme is the physically correct one is still unclear. For a free photon, the physically correct separation scheme should be unique. In this paper we shall address the problem of angular momentum decomposition of a free photon. Based on the physical characteristic of the motion of the photon, we shall propose such a physical separation scheme with the hope that it solves this long-standing problem.

In relativistic quantum mechanics of the photon, the state of a free photon can be represented by its coordinate representation wavefunction, which is just the vector potential of a free electromagnetic field $\vec{A}(\vec{r},t)$ in the transverse gauge $\vec{\nabla} \cdot \vec{A}=0$. For a free photon, the Maxwell equation satisfied by the vector potential can be regarded as the "Schr$\rm{\ddot{o}}$dinger eqution" of the photon:
\begin{equation}
\frac{\partial^2 \vec{A}}{\partial t^2}-\triangle \vec{A}=0,  ~~~\vec{\nabla}\cdot \vec{A}=0.
\end{equation}
According to the standard definition of quantum theory, the total angular momentum operator is the generator of spatial rotation of the system.
Under a spatial rotation $\vec{r}' = R\vec{r}$, the photon wavefunction transforms as a vector wavefunction: $\vec{A}'(\vec{r})= R \vec{A}(R^{-1}\vec{r})$. From this one can deduce the angular momentum operator for a free photon:
\begin{equation}
\hat{\vec{J}}=\hat{\vec{L}} +\hat{\vec{S}}
\end{equation}
with $\hat{\vec{L}}=\hat{\vec{r}}\times \hat{\vec{k}}$ being the usual orbital angular momentum operator and $\hat{\vec{S}}=(\hat{S}_x, \hat{S}_y,\hat{S}_z)$ being the usual spin operator for a spin-one particle:
\[
\hat{S}_x=\left(
          \begin{array}{ccc}
          0 & 0 & 0 \\
          0 & 0 & -i \\
          0 & i & 0
          \end{array}
           \right)
~~\hat{S}_y=\left(
          \begin{array}{ccc}
         0 & 0 & i \\
          0 & 0 & 0 \\
          -i & 0 & 0
         \end{array}
           \right)
~~\hat{S}_z=\left(
         \begin{array}{ccc}
          0 & -i & 0 \\
          i & 0 & 0 \\
          0 & 0 & 0
         \end{array}
         \right)
\]
For a massive spin-one particle, $\hat{\vec{L}}$ and $\hat{\vec{S}}$ can be naturally identified as the orbital angular momentum operator and spin operator, respectively. However, for a photon, such an identification fails because the actions of $\hat{\vec{L}}$ and $\hat{\vec{S}}$ do not preserve the transversality condition of the photon wavefunction \cite{CTDG} and hence they could not be regarded as operators acting on the physical state space of the photon.

Mathematically, a relativistic particle with definite mass and spin corresponds to a definite unitary representation of the Poincare
group. Here, we shall discuss the problem of separation of photon angular momentum from the point of view of representation theory of
the Poincare group. Since photon is a massless particle with spin one, according to representation theory of the Poincare group, a photon
with definite momentum $\vec{k}$ has two different helicity states which we denote as $|\vec{k},\lambda \rangle$ with $\lambda = \pm 1$
being the helicity quantum number, and the totality of these physical states forms a representation space of the Poincare group. On this representation space the three generators of spatial translation are the momentum operators of the photon and the three generators of rotation are the angular momentum operators of the photon.

Let us now turn to the problem of separation of photon angular momentum. Suppose there exist a definite and physically meaningful way to separate
the photon angular momentum operator into an orbital part and a spin part, then one can write
\begin{equation}
\hat{\vec{J}}=\hat{\vec{L}}+\hat{\vec{S}},
\end{equation}
where the orbital angular momentum operator $\hat{\vec{L}}$ and the spin operator $\hat{\vec{S}}$ are both operators acting on the physical state space of the photon (here, for clarity of notation, we use the same notations as the above to denote the orbital angular momentum operator and spin operator we propose to define). Now, the spin operator $\hat{\vec{S}}$, which describes the intrinsic angular momentum of the photon, should not depend on the specification of the origin of the coordinate system; hence it should commute with the generator of spatial translation, namely, the momentum operator: $[\hat{S}_i,\hat{k}_j]=0$.

Now, we shall show that the three operators $\hat{S}_x$, $\hat{S}_y$, $\hat{S}_z$, even though we call them spin operators, could not satisfy the usual angular momentum algebra. The proof is as follows. Suppose one could define the three spin operators $\hat{S}_x$, $\hat{S}_y$, $\hat{S}_z$
in such a way that $[\hat{S}_i,\hat{S}_j]=i \varepsilon_{ijk}\hat{S}_k$ is satisfied. Then, following the standard way of angular momentum theory, one could introduce the raising and lowering operators: $\hat{S}_{\pm} =\hat{S}_x\pm i \hat{S}_y$, which satisfy the following set of commutation relations:
\begin{eqnarray}
&& [\hat{S}_{z},\hat{S}_{+}] = \hat{S}_{+} \label{a}  \\
&&[\hat{S}_{z},\hat{S}_{-}] = -\hat{S}_{-} \label{b}  \\
&&[\hat{S}_{+},\hat{S}_{-}] = 2\hat{S}_{z}. \label{c}
\end{eqnarray}
The raising and lowering operators $\hat{S}_{\pm}$ also commute with the momentum operator. Now, let us consider a photon state with momentum
$\vec{k}=(0,0,k)$ along the z-direction and definite helicity $\lambda$. One then has
\begin{equation}
\hat{J}_z | \vec{k},\lambda \rangle=\lambda |\vec{k},\lambda \rangle.
\end{equation}
Now, according to the physical meaning of orbital angular momentum, the component of orbital angular momentum of the photon along its direction
of motion should vanish, thus $\hat{L}_z | \vec{k},\lambda \rangle=0$, and hence one has $\hat{S}_z |\vec{k}, \lambda \rangle =\lambda | \vec{k},\lambda \rangle$. Next, let us first consider the state $\hat{S}_{+} |\vec{k}, \lambda=+1 \rangle $. We first note that
\begin{equation}
\hat{\vec{k}}~ \hat{S}_{+} |\vec{k}, \lambda=+1 \rangle =\vec{k}~ \hat{S}_{+} |\vec{k}, \lambda=+1 \rangle,
\end{equation}
which shows that the state vector $\hat{S}_{+} |\vec{k}, \lambda=+1 \rangle$, if it is nonzero, also represents a one-photon state with momentum
$\vec{k}$. Then, from Eq. (\ref{a}) one can deduce
\begin{equation}
\hat{S}_z \hat{S}_{+} | \vec{k}, \lambda=+1 \rangle =2 \hat{S}_{+} |\vec{k}, \lambda=+1 \rangle.
\end{equation}
Thus, the state vector $\hat{S}_{+} |\vec{k}, \lambda=+1 \rangle$ should represent a one-photon state moving along the z-direction with helicity two. However, in the physical state space of the photon there is no such state, hence one necessarily has
$\hat{S}_{+} |\vec{k}, \lambda=+1 \rangle=0$. Then, let us consider the state $\hat{S}_{-} |\vec{k}, \lambda=+1 \rangle$. It can be easily seen that this state vector also represents a one-photon state with momentum $\vec{k}$. From Eq. (\ref{b}) one can deduce
\begin{equation}
\hat{S}_z \hat{S}_{-} | \vec{k},\lambda=+1 \rangle =0.
\end{equation}
Now, we show that the state vector $\hat{S}_{-} |\vec{k}, \lambda=+1 \rangle$ cannot be zero. First, from Eq. (\ref{c}) one can deduce
\begin{equation}\label{CommutatorRelation}
\hat{S}_{+} \hat{S}_{-} | \vec{k},\lambda=+1 \rangle -\hat{S}_{-} \hat{S}_{+} | \vec{k},\lambda=+1 \rangle =2 |\vec{k}, \lambda=+1 \rangle.
\end{equation}
One already has $\hat{S}_{+} |\vec{k}, \lambda=+1 \rangle=0$, so, if one also has $\hat{S}_{-} | \vec{k}, \lambda=+1 \rangle=0$, this will contradict Eq. (\ref{CommutatorRelation}).
Thus, the state vector $\hat{S}_{-} |\vec{k},\lambda=+1 \rangle $ is nonzero and it represents a one-photon state moving along the z-direction but with helicity zero. But we all know that in the physical state space of the photon there is no such state! Clearly, the appearance of this contradictory conclusion is due to our presupposition that the three photon spin operators $\hat{S}_x,\hat{S}_y, \hat{S}_z$ satisfy the angular momentum algebra. Therefore, we conclude that in the physical state space of the photon one cannot define a set of photon spin operators satisfying the angular momentum algebra.

Then, how should one look upon the concept of photon spin? The usual definition of the spin of a particle as the angular momentum of this particle in its rest frame does not apply to the photon since a photon always moves with the speed of light. For a photon there always exists a distinctive direction in space---the direction of its momentum vector. In this case, the system does not have the symmetry under the whole three-dimensional rotation group, but only the axial symmetry with respect to that specific axis. Under the condition of axial symmetry, the conserved quantity is the helicity of the photon---the component of the angular momentum along its direction of motion, which also coincides with the component of the spin angular momentum along the direction of motion \cite{BLP}. Thus, based on such a physical characteristic of the motion of the photon, we propose the
following physical definition for the spin of a photon: a photon with definite momentum $\vec{k}$ and helicity $\lambda$ has a definite spin vector $\vec{s}=\lambda \frac{\vec{k}}{|\vec{k}|}$. Based on this definition, the action of the photon spin operator (which we shall denote as $\hat{\vec{S}}_{phys}$) on a photon state with definite momentum and helicity can be expressed as
\begin{equation}
\hat{S}_{phys,i} |\vec{k},\lambda \rangle =\lambda \frac{k_i}{|\vec{k}|} |\vec{k},\lambda \rangle, ~~~(i=x,y,z).
\end{equation}
The three components of the photon spin operator thus defined commute with each other. This could be verified by direct calculation:
\begin{eqnarray}
[\hat{S}_{phys,i},\hat{S}_{phys,j}]|\vec{k},\lambda \rangle &=& \frac{k_i k_j}{|\vec{k}|^2} |\vec{k},\lambda \rangle-\frac{k_j k_i}{|\vec{k}|^2} |\vec{k},\lambda \rangle \nonumber \\
&=& 0.
\end{eqnarray}
One can also verify that each component of $\hat{\vec{S}}_{phys}$ commutes with the momentum operator:
\begin{eqnarray}
[\hat{S}_{phys,i},\hat{k}_j]|\vec{k},\lambda \rangle &=& \lambda \frac{k_i k_j}{|\vec{k}|} |\vec{k},\lambda \rangle -\lambda \frac{k_j k_i}{|\vec{k}|} |\vec{k},\lambda \rangle \nonumber \\
&=& 0,
\end{eqnarray}
which agrees with the usual concept of the spin angular momentum of a particle.

Now that we have given a physical definition for the photon spin operator, we can give an explicit expression for this operator using the photon wavefunction as a concrete realization of the photon state. The wavefunction for a photon with definite momentum $\vec{k}$ and helicity $\lambda$
can be written as
\begin{equation}
\vec{A}^{(\lambda)}_{\vec{k}} (\vec{r})=\frac{1}{\sqrt{2\omega}}\vec{e}^{~(\lambda)}_{\vec{k}} e^{i \vec{k}\cdot \vec{r}}, ~~\omega=|\vec{k}|,
\end{equation}
where $\vec{e}^{~(\lambda)}_{\vec{k}}$ denotes the polarization vector of the photon with $\lambda=+1$ corresponding to the right-handed circular
polarization and $\lambda=-1$ corresponding to the left-handed circular polarization. The polarization vectors have the following explicit expressions:
\begin{equation}
\vec{e}^{~(+1)}_{\vec{k}}=-\frac{1}{\sqrt{2}}(\vec{e}_\xi+i\vec{e}_\eta),~~~\vec{e}^{~(-1)}_{\vec{k}}=\frac{1}{\sqrt{2}}(\vec{e}_\xi-i\vec{e}_\eta),
\end{equation}
where $(\vec{e}_\xi,\vec{e}_\eta, \frac{\vec{k}}{|\vec{k}|})$ forms a right-handed orthogonal dreibein. Now, since one has $\hat{\vec{S}}\cdot \frac{\vec{k}}{|\vec{k}|} ~\vec{e}^{~(\lambda)}_{\vec{k}}= \lambda ~\vec{e}^{~(\lambda)}_{\vec{k}}$, one obtains
\begin{equation}
\hat{\vec{S}}\cdot \frac{\vec{k}}{|\vec{k}|}~ \vec{A}^{(\lambda)}_{\vec{k}} (\vec{r})=\lambda \vec{A}^{(\lambda)}_{\vec{k}} (\vec{r})
\end{equation}
and
\begin{eqnarray}
\hat{\vec{J}}\cdot \frac{\vec{k}}{|\vec{k}|} ~ \vec{A}^{(\lambda)}_{\vec{k}} (\vec{r})&=& \hat{\vec{L}} \cdot \frac{\vec{k}}{|\vec{k}|}~  \vec{A}^{(\lambda)}_{\vec{k}} (\vec{r})+ \hat{\vec{S}}\cdot \frac{\vec{k}}{|\vec{k}|}~ \vec{A}^{(\lambda)}_{\vec{k}} (\vec{r}) \nonumber \\
&=& \lambda \vec{A}^{(\lambda)}_{\vec{k}} (\vec{r}).
\end{eqnarray}
Therefore, when one uses the coordinate representation wavefunction to represent a photon state, the photon spin operator $\hat{\vec{S}}_{phys}$ has the following explicit expression:
\begin{equation}\label{S-expression}
\hat{\vec{S}}_{phys}=\hat{\vec{S}}\cdot \frac{\hat{\vec{k}}}{|\hat{\vec{k}}|}\frac{\hat{\vec{k}}}{|\hat{\vec{k}}|}.
\end{equation}
The orbital angular momentum operator of the photon is then naturally defined according to the separation $\hat{\vec{J}}=\hat{\vec{L}}_{phys}+\hat{\vec{S}}_{phys}$:
\begin{equation}\label{L-expression}
\hat{\vec{L}}_{phys}=\hat{\vec{L}}+\hat{\vec{S}}-\hat{\vec{S}}\cdot \frac{\hat{\vec{k}}}{|\hat{\vec{k}}|}\frac{\hat{\vec{k}}}{|\hat{\vec{k}}|}.
\end{equation}
It can be easily verified that $\hat{\vec{L}}_{phys} \cdot \frac{\hat{\vec{k}}}{|\hat{\vec{k}}|}=0$, which agrees with the usual concept of orbital
angular momentum of a particle. In his discussion on Einstein-Podolsky-Rosen-Bohm experiment with relativistic massive
particles, Czachor suggested that the orbital angular momentum and spin operators defined via the relativistic center-of-mass operator should be the most physical definition of orbital angular momentum and spin for a relativistic particle \cite{Czachor}. For a massless photon the orbital angular momentum and spin operators suggested by Czachor reduce to the ones given in Eqs. (\ref{S-expression}) and (\ref{L-expression}). We note that the coordinate representation operators given in Eqs. (\ref{S-expression}) and (\ref{L-expression}) were also obtained by Bliokh {\it et al.} \cite{BAOA} in the context of discussing angular momenta and spin-orbit interaction of nonparaxial light in free space.

As discussed above, the three components of $\hat{\vec{S}}_{phys}$ commutes with each other: $[\hat{S}_{phys,i}, \hat{S}_{phys,j}]=0$. Now, by construction both $\hat{\vec{S}}_{phys}$ and $\hat{\vec{L}}_{phys}$ are vector operators, hence one has
\begin{eqnarray}
&&[\hat{J}_i,\hat{S}_{phys,j}]= i \varepsilon_{ijk}\hat{S}_{phys,k} \nonumber \\
&&[\hat{J}_i,\hat{L}_{phys,j}]= i \varepsilon_{ijk}\hat{L}_{phys,k}.
\end{eqnarray}
The above two relations together with $[\hat{S}_{phys,i}, \hat{S}_{phys,j}]=0$ imply
\begin{eqnarray}
&&[\hat{L}_{phys,i},\hat{S}_{phys,j}]=i\varepsilon_{ijk} \hat{S}_{phys,k} \label{LS},  \\
&&[\hat{L}_{phys,i},\hat{L}_{phys,j}]=i\varepsilon_{ijk} (\hat{L}_{phys,k}-\hat{S}_{phys,k}). \label{LL}
\end{eqnarray}
Now, since the three operators $\hat{S}_{phys,i} (i=x,y,z)$ commute with each other, they are not angular momentum operators in the usual sense.
Nevertheless, the component of $\hat{\vec{S}}_{phys}$ on the direction of motion of the photon generates spin rotation of the polarization of the photon:
\begin{equation}\label{S}
e^{-i \theta \hat{\vec{S}}_{phys}\cdot \frac{\vec{k}}{|\vec{k}|}} |\vec{k},\lambda \rangle=e^{-i \lambda \theta}|\vec{k},\lambda \rangle.
\end{equation}
The three operators $\hat{L}_{phys,i}(i=x,y,z)$ also do not satisfy the angular momentum algebra. They even do not form a set of generators of some Lie algebra. Hence, they are also not angular momentum operators in the usual sense. From Eq. (\ref{LS}) it is seen that different components of
$\hat{\vec{L}}_{phys}$ and $\hat{\vec{S}}_{phys}$ do not commute. However, if one splits the component of the total angular momentum operator along
an arbitrary direction $\vec{n}$ into the orbital part and spin part:
\begin{equation}
\hat{\vec{J}} \cdot \vec{n}=\hat{\vec{L}}_{phys} \cdot \vec{n}+\hat{\vec{S}}_{phys}\cdot \vec{n},
\end{equation}
then it can be easily seen that the orbital part and the spin part commute:
\begin{equation}
[\hat{\vec{L}}_{phys} \cdot \vec{n},~\hat{\vec{S}}_{phys}\cdot \vec{n}]=0.
\end{equation}
On the other hand, one has $\hat{\vec{L}}_{phys} \cdot \frac{\vec{k}}{|\vec{k}|} ~|\vec{k},\lambda \rangle=0$, which implies
\begin{equation}\label{L}
e^{-i \theta \hat{\vec{L}}_{phys} \cdot \frac{\vec{k}}{|\vec{k}|}} |\vec{k},\lambda \rangle = |\vec{k},\lambda \rangle.
\end{equation}
Combining Eqs. (\ref{S}) and (\ref{L}), one has
\begin{eqnarray}
e^{-i \theta \hat{\vec{J}}\cdot \frac{\vec{k}}{|\vec{k}|}} |\vec{k}, \lambda \rangle &=&  e^{-i\theta \hat{\vec{L}}_{phys} \cdot \frac{\vec{k}}{|\vec{k}|} } ~ e^{-i \theta \hat{\vec{S}}_{phys}\cdot \frac{\vec{k}}{|\vec{k}|}} |\vec{k}, \lambda \rangle \nonumber \\
&=& e^{-i \lambda \theta }|\vec{k}, \lambda \rangle,
\end{eqnarray}
which is in perfect accordance with the physical meaning of $\hat{\vec{J}}\cdot \frac{\vec{k}}{|\vec{k}|}$ as the helicity operator of the photon.

Thus, based on the physical characteristic of the motion of the photon, we could obtain a natural and physical decomposition of the total angular
momentum operator for a free photon into an orbital part and a spin part. We hope that this solves the long-standing problem of angular momentum decomposition of a free photon and it could provide a natural basis for further study of the problem of angular momentum decomposition of a gauge field system.

\end{document}